# Thermoelectricity in ternary rare-earth systems

V.N.Nikiforov[a], V.Yu.Irkhin[b], A.V. Morozkin[c]

[a]*Department of Physics, Moscow State University, Leninskie Gory, Moscow, GSP-2, 119899, Russia*
[b]*Institute of Metal Physics, Ekaterinburg, Russia*
[c]*Department of Chemistry, Moscow State University*

Crystallographic data, Seebeck coefficient, electrical resistance and thermal conductivity are reported for a large number of rare-earth compounds, manifestations of the Kondo effect being discussed. In more detail, thermoelectric properties of $Yb_3Co_4Ge_{13}$, $Yb_3Co_4Sn_{13}$ compounds and $Yb_2CeCo_4Ge_{13}$ and $Yb_{2.3}La_{0.7}Co_4Ge_{13}$ solid solutions are presented.

---

## I. Introduction

Thermoelectric materials are of great interest due to their ability to use waste heat to generate electricity and act as solid state Peltier coolers. Thermoelectric devices based on thermoelectric materials have numerous advantages, such as being low-noise, high reliability, without any moving partsand long period of operation. Since power units of thermoelectric generation produce electric current from temperature difference, we can convert waste heat energy to electric energy. Therefore, thermoelectric power generation is one of the most promising tools for environmental conservation. However, the conversion efficiency of conventional thermoelectric generation is insufficient in performance. The reason for this low efficiency is in poor thermoelectric properties of most conventional thermoelectric materials.

The thermoelectric fitness of a material is estimated by the value of the dimensionless parameter, figure of merit $ZT = S^2\sigma T/\kappa$ where $S$ is the Seebeck coefficient, σ electrical conductivity, and κ thermal conductivity. The material having higher $ZT$ value possesses better thermoelectric performance. High-performance thermoelectric materials must have large $S$, high σ, and low κ to retain the heat at the junction and reduce the heat transfer losses.

*How to increase ZT?*

The Wiedemann–Franz law limits the ratio of the electronic contribution to the thermal conductivity and electrical conductivity of a metal. That ratio should be proportional to the temperature, $\kappa/\sigma = LT$. Rosenberg (2004) notes that the Wiedemann–Franz law is generally valid for high and low (i.e., a few of Kelvins) temperatures, but can be violated at intermediate temperatures. In degenerate semiconductors, the Lorentz number $L$ has a strong dependence on system parameters: dimensionality, strength of interatomic interactions and Fermi level position. The Lorentz number can be reduced by manipulating electronic density of states, varying doping density and layer thickness in superlattices and composite materials with correlated carriers (Minnich et al 2009, Pichanusakorn and Bandaru 2009).

We can use the following strategies:

- To reduce thermal conductivity and increase the efficiency by minimizing the thermal losses through the device.
- To increase the Seebeck coefficient and the electrical conductivity. Recent advances that achieved $ZT \sim 2$ are mostly due to the reduction of χ by nanostructuring, i.e., by using multilayered materials (Kanatzidis 2010) and nanocomposites (Minnich et al 2009)
- To introduce resonant states within the Fermi window by doping (Mahan and Sofo 1996)

## II. Skutterudite systems

Skutterudite compounds $MX_3$ with a cubic structure (M = Co, Rh, Ir, Fe, Ru, X = P, As, Sb) are a class of promising high performance thermoelectric material for power generation (Morelli et al1995, 1997; Caillat et al 1996; Nolas et al 1996) and thus have attracted a great deal of interest in recent years.

Although binary skutterudite compounds possess good electrical transport properties, their overall figure of merit *ZT* is not too high because of relatively high thermal conductivity. It is found by Sales et al (1996) and Nolas et al (1996a, 1998) that rare-earth atoms such as La, Ce, etc. may be inserted into the voids of the crystal structure where they "rattle" around their equilibrium positions. This rattling motion can efficiently scatter the phonons and thus greatly reduce the lattice thermal conductivity without deteriorating the electrical transport properties. The *ZT* value of the rare-earth filled skutterudite is more than unity between 773 and 973 K (Sales et al 1996).

Ternary skutterudites $R_yM_4X_{12}$; with R being rare earth, M transition metal and X pnictogen; are a novel class of materials exhibiting extraordinary large thermoelectric potential, Kondo- and heavy-fermion properties (Nolas et al 1999, Bauer et al 2003).

The interest in skutterudite-related systems is connected with search of new thermoelectric materials (Rowe 2006). Unfilled skutterudites of the type $M(P,Sb,As)_3$ (M is a transition metal) contain voids into which low-coordination ions, in particular rare earth elements, can be inserted. This increases phonon scattering and decreases lattice thermal conductivity without increasing electrical resistitivity $\rho$ (Nolas et al 1996a). Thus the figure of merit can become rather large.

The electronic properties of the $Ir_4LaGe_3Sb_9$, $Ir_4NdGe_3Sb_9$, and $Ir_4SmGe_3Sb_9$ systems were investigated by Nolas et al (1996a). Another class of skutterudite-related systems are $R_3M_4X_{13}$ compounds (where R is a rare earth element, X = Ge or Sn) which attract attention due to the interesting electron properties, interplay of superconductivity and magnetic order (Mudryk et al 2001). In particular, the Yb-based systems demonstrate the intermediate valence nature of the Yb ions and slightly enhanced value of the linear electronic specific heat coefficient.

## III. Theoretical sketch for thermoelectric power in rare-earth and Kondo systems

The Seebeck coefficient can become very large due to the reduction of free carriers; then there is a crossover from positive to negative thermopower as soon as electron transport dominates over hole transport.

The Kondo skutterudites may exhibit large *S* values due to an appropriate reduction of free charge carriers caused by rare earth elements like Ce or Yb (Bauer et al 2003). Strong electron correlations, as obvious in $Pr_{0.73}Fe_4Sb_{12}$; can enhance significantly thermopower values above those found in skutterudites containing valence-stable rare earth elements (Nolas et al 1999).

Based on the *s-d* interaction model for dilute magnetic alloys Kondo (1969) has calculated the scattering probability of the conduction electrons to the second Born approximation. Thus the effect gives rise to a singular term in the resistivity which involves a factor of $cJ^3\ln T$, where $c$ is the concentration of impurity atoms, $J$ the *s-d* exchange integral. When combined with the lattice resistivity, this gives rise to a resistance minimum, provided $J$ is negative, and large temperature-independent Seebeck coefficient (Kondo 1965, 1969).

Here we discuss briefly the Seebeck coefficient $S(T)$ in Kondo lattices following to Irkhin and Katsnelson (1989), Irkhin and Irkhin (2007). We can distinguish two cases:

(i) *Perturbation theory regime.* The Kondo correction in the electron self-energy $\Sigma(E)$ (or conduction-electron T-matrix) is proportional to $J^3\ln(E)$. Large Kondo contributions to $S(T)$ correspond to the anomalous odd contribution to the relaxation rate $\tau^{-1}(E)$ (Kondo 1969). The latter should arise, by analytical properties of $\Sigma(E)$, from the logarithmic singularity in $\mathrm{Re}\Sigma(E)$ (Irkhin and Katsnelson 1989). Although such a singularity is absent in the usual Kondo problem, it occurs in the presence of the potential scattering $V$ which leads to emergence of complex factors $1 + V\Sigma_k(E - t_k + i0)^{-1}$ which "mix" $\mathrm{Im}\Sigma$ and $\mathrm{Re}\Sigma$ in the incoherent regime.

In the presence of interaction between magnetic moments (Kondo lattices), spin dynamics leads to the replacements $\ln|E| \rightarrow (1/2)\ln(E^2 + \omega^2)$, $\mathrm{sign}E \rightarrow (2/\pi)\tan^{-1}(E/\omega)$ in ImΣ and ReΣ, respectively (ω is a characteristic spin-fluctuation frequency). Then the anomalous contribution reads $S(T) \sim e\rho(T) T/\max\{T,\omega\}$. Thus the quantity ω plays the role of a characteristic fluctuating magnetic field which is introduced by Kondo (1969) to describe thermoelectric power of diluted Kondo systems.

In real concentrated systems Kondo systems, at moderately high (as compared to $T_K$) temperatures $S(T)$ is usually large and has an extremum (a maximum at $S < 0$, a minimum at $S < 0$).

(ii) *The low-temperature regime.* Besides the Kondo temperature, one introduces the second energy scale - the coherence temperature $T_{coh}$, which corresponds to onset of coherent Kondo scattering by different lattice sites. This is usually small in comparison with $T_K$. The picture of the coherent state formation enables one to treat experimental data on low-temperature anomalies of thermoelectric power in heavy-fermion systems. With decreasing $T$ below the above-discussed high-temperature extremum, $S(T)$ often changes its sign, has an extremum again and vanishes linearly at $T \rightarrow 0$. Such a behaviour may be attributed to occurrence of a pseudogap with reversing the sign of the quantity $dN(E)/dE$ ($N(E)$ is the electron density of states) at the Fermi level, which determines the $S(T)$ sign in the Mott diffusion mechanism.

Generally, the systems having low characteristic temperature scales exhibit enhanced $S$ values. In particular, heavy-fermion systems, which exhibit sharp features in density of states due to the hybridization between conduction electrons and f or d-electrons, have been considered as candidates for low-temperature thermoelectric cooling applications (Zlatic et al 2009, Mahan 1998, Grenzebach 2006). Large thermopower values have been predicted and observed in Kondo insulators like $Ce_3Pt_3Sb_4$ at low temperatures (Mahan et al 1997, Takabatake et al 2003).

In nanowire structures, electron transport is hardly affected by the boundary scattering due to their small intrinsic mean free paths while phonons are strongly scattered due to classical size effect. These results suggest that the nanostructures of Kondo insulators can be designed for high performance thermoelectric cooling devices at low temperatures.

## IV. Experimental results

We have synthesized and investigated a large number of binary and ternary rare-earth compounds and alloys, including new ones (Tables 1-4). Here we pay attention to Ce-based Kondo compounds. Maximum *ZT* values for the presented systems (which are shown bold) are about 7%. One can see that Kondo-lattice systems like $CeSi_2$, $Ce_{1-x}Y_xNiSb$ indeed have largest Seebeck coefficient values. Moreover, such systems demonstrate strong temperature dependences of resistivity and Seebeck coefficient (Fig.1).

We report also in more details the results on thermoelectric properties of the scutterudites $Yb_3Co_4Sn_{13}$ and $Yb_3Co_4Ge_{13}$ compounds and $Yb_{3-x}R_xCo_4Ge_{13}$ (R = Ce, La) solid solutions (Nikiforov et al 2013).

Analysis of the powder X-ray diffractograms shows that the $R_3T_4X_{13}$ compounds (R = rare earth, T = Ru, Rh, Os, X = Ge, Sn) adopt the $Pr_3Rh_4Sn_{13}$-type cage-like structure. This favours small lattice thermal conductivity.

The electrical resistance along the long axis of the sample was measured with a standard four-terminal geometry using a dc current of 0.8 mA The Seebeck coefficient (the thermoelectric voltage under zero electric current) has been measured with disconnection of the sample from the current source. The heat flux in the sample was generated with electric heater. The thermovoltage $E$ was measured over the same potential contacts used in the resistance measurements. The Seebeck coefficient was calculated as the ratio $E$/dT. The uncertainty of $S$ was estimated to be less than 10%. The thermal conductivity was measured by a longitudinal steady state method.

All these compounds demonstrate a metallic-type conductivity. Probably, the $Yb_3Co_4Ge_{13}$ has higher electric resistance due to presence of Ge semiconductor in the alloy: the resistance of the $Yb_2CeCo_4Ge_{13}$ and $Yb_{2.3}La_{0.7}Co_4Ge_{13}$ alloys is less than of $Yb_3Co_4Ge_{13}$ alloy which demonstrates absence of the Ge admixture phase.

The Seebeck coefficients, electrical resistance and thermal conductivity have monotonic behavior with increasing temperature from 240 to 380 K (Fig. 2). One can see that the lattice thermal conductivity $\kappa_{lattice}$ of Ge-containing $Pr_3Rh_4Sn_{13}$-type compounds is higher than for Sn-containing compound (thermal coductivity κ is the sum of lattice and electronic conductivity, $\kappa = \kappa_{lat} + \kappa_{el}$) due to that the mass of Sn is higher than the mass of Ge. Also, due to this, the Wiedemann-Frantz parameter $WF = \pi\kappa/\gamma_{WF}T$ has higher values for Ge-containing compounds (Nikiforov et al 2013).

The Seebeck coefficient is $S$ = 14 ? 27 µV//K for $Yb_3Co_4Sn_{13}$, and $S$ = -21 ? -12 µV/K for $Yb_3Co_4Ge_{13}$. Among the scutterudite systems under consideration, the parameter $ZT$ is maximal for $Yb_3Co_4Sn_{13}$ compounds and acquires the value of 0.017 at 380 K. Although itsthermal conductivity is not maximal, this compound demonstrates lowest resistivity at high temperatures. Thus a good metallic electrical resistivity and high Seebeck coefficient (rather than small heat conductivity) turn out to be decisive for large $ZT$ values in this case. Provided that the dependences of electric resistance, thermal conductivity and Seebeck coefficient vs. temperature retain for $Yb_3Co_4Sn_{13}$ compound up to 1000 K, this compound may have the $ZT$ value about 0.3 at 1000 K.

Substitution of Yb for Ce or La, as well as substitution of Ge for Sn shifts the Seebeck coefficient to positive values observed in the $Yb_3Co_4Ge_{13}$ compound. The Seebeck coefficient of $Yb_{2.3}La_{0.7}Co_4Ge_{13}$ demonstrates non-monotonous temperature dependence. This sign change may be connected with a non-trivial electronic structure near the Fermi level, characteristic for Kondo-like and intermediate-valence systems.

The parameter $ZT = TS^2/(\rho\ \kappa)$ is maximal for $Yb_3Co_4Sn_{13}$ compounds and acquires the value of 0.017 at 380 K (Fig. 2). Although its κ is not maximal, this compound demonstrates lowest resistivity at high temperatures. Thus a good metallic electrical resistivity and high Seebeck coefficient (rather than small heat conductivity) turn out to be decisive for large $ZT$ values.

We can conclude as follows:

(i) The $Pr_3Rh_4Sn_{13}$-type Ge-containing compounds have lattice thermal conductivity higher than Sn-containing compounds due to relatively low Ge atom mass as compared to Sn, which is important for thermoelectric characteristics.

(ii) The systems investigated demonstrate a non-trivial behavior vs. alloy composition. Substitution of Yb for cerium or La shifts the Seebeck coefficient to positive values. A more systematic treatment is of interest. Probably, some $Yb_{3-x}R_xCo_4Sn_{13}$ compounds will have the Seebeck coefficient and $ZT$ parameter value higher than $Yb_3Co_4Sn_{13}$ compound.

Although $ZT$ values obtained in the rare-earth compounds are still not too large, they are promising materials. There exist some ways to improve their characteristics. Further experimental investigations in this directions and comparison with other ternary rare-earth systems would be useful.

**Acknowledgments**

This work was supported by Asahi Kasei Corporation (Japan) in the ISTC project N 2382p and by the Programs of fundamental research of RAS Physical Division “Strongly correlated electrons in solids and structures”, project No. 12-T-2-1001 (Ural Branch) and of RAS Presidium “Quantum mesoscopic and disordered structures”, project No. 12-P-2-1041.

**Caption to figures**

Fig.1. Resistivity and Seebeck coefficient vs. temperature for $CeSi_2$ (a) and $Ce_{0.5}Y_{0.5}CoSi_3$ (b) alloys

Fig.2. Seebeck coefficient *S*, electric resistance ρ, thermal conductivity κ and *ZT* parameter vs. temperature for $Yb_3Co_4Ge_{13}$, $Yb_2CeCo_4Ge_{13}$, $Yb_{2.3}La_{0.7}Co_4Ge_{13}$ and $Yb_3Co_4Sn_{13}$ compounds.

Table 1

Composition, cell parameters *a, b* and *c* (nm) of silicides and their solid solutions. Physical properties: electrical resistivity ρ (μOhm•m), Seebeck coefficient *S* (μV/K), thermal conductivity κ (W/m•K) and *ZT* parameter at 300 K. The Wiedemann-Franz parameter is WF = $\rho \bullet \kappa/\gamma_{WF} \bullet 300$, $\gamma_{WF} = (\pi^2/3) \bullet (k_B/e)^2$ ($k_B$ is the Boltzmann constant and *e* is the electronic charge)

| **Alloys**<br>(sample) | **Mass fraction** | **Phase** | **Structure** | **a** | **b** | **c** | **S** | ρ | κ | **WF** | **ZT** |
|---|---|---|---|---|---|---|---|---|---|---|---|
| $CeCoSi_3$ | 0.80<br>0.07<br>0.13 | $Ce_{20}Co_{19}Si_{60}$<br>$Co_{33}Si_{66}$<br>$Ce_{25}Co_{19}Si_{56}$ | $BaNiSn_3$<br>$CaF_2$<br>$CeNiSi_2$ | 0.41218(6)<br>0.5348(9)<br>0.4122(3) | <br><br>1.6391(9) | 0.9534(2)<br><br>0.4139(3) | -34.6 | 2.1 | 9.1 | 2.61 | 0.018 |
| $CeCo_{0.8}Ni_{0.2}Si_3$ | 0.73<br>0.08<br>0.19 | **$Ce_{21}Co_{16}Ni_4Si_{59}$**<br>$Co_{29}Ni_4Si_{67}$<br>$Ce_{25}Co_{15}Ni_5Si_{54}$ | $BaNiSn_3$<br>$CaF_2$<br>$CeNiSi_2$ | 0.41391(5)<br>0.5355(6)<br>0.4149(2) | <br><br>1.6465(1) | 0.9531(1)<br><br>0.4070(1) | -37.5 | 1.5 | 16.6 | 3.4 | 0.017 |
| $CeCo_{0.8}Cu_{0.2}Si_3$ | 0.90<br>0.03<br>0.05 | **$Ce_{22}Co_{15}Cu_3Si_{59}$**<br>$Ce_{33}Co_2Cu_9Si_{56}$<br>$Si_{97}$<br>$Ce_{21}Co_3Cu_{33}Si_{43}$ | $BaNiSn_3$<br>$CeNiSi_2$<br>C<br>$CeGa_2Al_2$ | 0.41307(4)<br>0.4145(4)<br>0.5416(8) | <br>1.6337(13) | 0.9545(1)<br>0.4175(4) | -36.9 | 0.91 | 18.8 | 2.33 | 0.024 |
| $CeCoSi_3$ | 0.73<br>0.10<br>0.17 | $CeCoSi_3$<br>$CoSi_2$<br>$CeCoSi_2$ | $BaNiSn_3$<br>$CaF_2$<br>$CeNiSi_2$ | 0.4130(1)<br>0.5355(5)<br>0.4121(3) | <br><br>1.6480(9) | 0.9545(2)<br><br>0.4119(4) | -35.2 | 2.27 | 20.2 | 6.26 | $8.1 \bullet 10^{-3}$ |
| $CeCoSi_{1.5}Ge_{1.5}$ | 0.98<br>0.02 | **$CeCoSi_{1.5}Ge_{1.5}$**<br>$CeSi_{2-x}Ge_2$ | $BaNiSn_3$<br>$ThSi_2$ | 0.4220(1)<br>0.4258(4) | | 0.9686(1)<br>1.402(1) | 25.74 | 1.11 | 10.9 | 1.65 | 0.0164 |
| $Ce_{0.5}Y_{0.5}CoSi_3$ | 0.78<br>0.05<br>0.14<br>0.03 | **$Y_{10}Ce_{10}Co_{20}Si_{60}$**<br>$Co_{33}Si_{66}$<br>$Y_{13}Ce_{12}Co_{20}Si_{55}$<br>$Si_{95}$ | $BaNiSn_3$<br>$CaF_2$<br>$CeNiSi_2$<br>C | 0.40947(4)<br>0.5358(7)<br>0.4065(2)<br>0.5424(9) | <br><br>1.6402(6) | 0.9538(1)<br><br>0.4030(1) | -30.8 | 3.8 | 15.5 | 8.04 | $4.8 \bullet 10^{-3}$ |
| $Ce_{0.5}Y_{0.5}CoSi_3^{HP}$ | 0.94<br>0.04 | **$Ce_{10}Y_{12}Co_{18}Si_{60}$**<br>$Co_{32}Si_{68}$<br>$Si_{95}$ | $BaNiSn_3$<br>$CaF_2$<br>C | 0.40897(3)<br>0.5348(6) | | 0.95189(8) | -37.6 | 0.45 | 16.4 | 1.01 | **0.057** |
| CeSiGe | 1.00 | **$Ce_{41}Si_{27}Ge_{32}$**<br>$S_{70}Ge_{30}$ | $ThSi_2$<br>C | 0.4243(1) | | 1.3897(3) | -3.4 | $5.9^{SM}$ | 8.0 | 6.44 | $7.4 \bullet 10^{-5}$ |
| $CeSi_{1.6}Ge_{0.4}$ | 1.00 | **$CeSi_{1.6}Ge_{0.4}$**<br>Si | $ThSi_2$<br>C | 0.4218(1) | | 1.3865(2) | -20.0 | $2.0^{SC}$ | 14.0 | 3.8 | $4.3 \bullet 10^{-3}$ |
| CeSiGa | 0.92<br>0.08 | **$Ce_{37}Si_{28}Ga_{35}$**<br>$Ce_{55}Si_{39}Ga_7$ | $ThSi_2$<br>FeB | 0.4229(1)<br>0.8277(5) | <br>0.3984(2) | 1.4353(2)<br>0.5972(4) | -11.4 | 1.02 | 14.4 | 2.00 | $2.6 \bullet 10^{-3}$ |
| $CeSi_2^{HP}$ | | $Ce_{34}Si_{67}$ | $ThSi_2$ | 0.41711(5) | | 1.3964(2) | -55.0 | 0.98 | 13.8 | 1.85 | **0.067** |
| $Ce_{0.5}Y_{0.5}Si_2^{HP}$ | 0.95<br>0.05 | **$Ce_{18}Y_{16}Si_{66}$**<br>$Si_{99}$ | $ThSi_2$<br>C | 0.41219(4) | | 1.3554(1) | -20.0 | 2.5 | 13.0 | 4.43 | $3.7 \bullet 10^{-3}$ |

$^{SC}$Semiconductor-type conductivity

$^{\mathbf{HP}}$High-purity sample

Table 2

Composition, cell parameters and physical properties of $NaZn_{13}$-type solid solutions

| **Alloys** (sample) | **Mass fraction** | **Phase** | **Structure** | **a** | **c** | **S** | **ρ** | **κ** | **WF** | **ZT** |
|---|---|---|---|---|---|---|---|---|---|---|
| $Ce_{0.2}La_{0.8}Co_{12}Si$<br>(3Q/N93 LCCI13) | 0.89<br>0.11 | $Ce_1La_8Co_{84}Si_8$<br>$Ce_2Co_{17}$ | $NaZn_{13}$<br>$Th_2Zn_{17}$ | 1.1285(2)<br>0.8350(9) | <br>1.221(1) | 15.0 | 0.64 | 21.3 | 1.86 | $4.95 \bullet 10^{-3}$ |
| $LaCo_{10}Ni_2Si$<br>(5Q/N26 LCNI13) | 0.91<br>0.05<br>0.04 | **$La_7Co_{72}Ni_{13}Si_9$**<br>$La_{15}Co_{42}Ni_{36}Si_8$<br>$Co_{87}Ni_{10}$ | $NaZn_{13}$<br>$CaCu_5$<br>Mg | 1.1292(2)<br>0.5028(4)<br>0.2507 | <br>0.3987(3)<br>0.4070 | 15.0 | $0.16^{SC*}$ | 16.2 | 0.35 | 0.026 |
| $LaCo_{10}Ni_2Si$<br>(5Q/N26 LCNI13) | 0.91<br>0.05<br>0.04 | **$La_7Co_{72}Ni_{13}Si_9$**<br>$La_{15}Co_{42}Ni_{36}Si_8$<br>$Co_{87}Ni_{10}$ | $NaZn_{13}$<br>$CaCu_5$<br>Mg | 1.1292(2)<br>0.5028(4)<br>0.2507 | <br>0.3987(3)<br>0.4070 | 14.2 | 0.67 | 18.8 | 1.72 | $4.8 \bullet 10^{-3}$ |
| $LaCo_{10}Fe_2Si$<br>(6Q/N48 LCFI13) | 0.97<br>0.02 | **$La_8Co_{76}Fe_{13}Si_9$**<br>$Co_{68}Fe_{32}$<br>$La_{25}Co_{44}Fe_5Si_{26}$ | $NaZn_{13}$<br>W<br>x | 1.1350(1)<br>0.2835(4) | | 14.0 | 0.69 | 18.9 | 1.78 | $4.5 \bullet 10^{-3}$ |
| $LaCo_{10}Cu_2Si$<br>(6Q/N58 LCPI13) | 0.65<br>0.26<br>0.09 | **$La_7Co_{76}Cu_5Si_{11}$**<br>$Co_{94}$<br>$La_{16}Co_{20}Cu_{64}$ | $NaZn_{13}$<br>Mg<br>$CaCu_5$ | 1.1290(6)<br>0.2500(3) | <br>0.4066(6) | 14.0 | 0.602 | 19.6 | 1.61 | $5.0 \bullet 10^{-3}$ |
| $Ce_{0.05}La_{0.95}Co_{10}Ni_2Si$<br>(6Q/N55 CLCNI13D) | 0.91<br>0.05<br>0.04 | **$Ce_{0.2}La_7Co_{71}Ni_{14}Si_8$**<br>$Ce_3La_9Co_{59}Ni_{23}Si_8$<br>$Co_{99}$ | $NaZn_{13}$<br>$CaCu_5$<br>Mg | 1.1290(2)<br>0.4966(5)<br>0.2507(9) | <br>0.4006(4)<br>0.4069(10) | 14.2 | 0.725 | 17.6 | 1.74 | $4.7 \bullet 10^{-3}$ |
| $Y_{0.05}La_{0.95}Co_{10}Ni_2Si$<br>(6Q/N56 YLCNI13) | 0.93 | **$Y_{0.2}La_7Co_{71}Ni_{14}Si_8$**<br>$Y_5La_7Co_{59}Ni_{23}Si_7$ | $NaZn_{13}$<br>x | 1.1301(2) | | 12.8 | 0.72 | 17.2 | 1.69 | $4.0 \bullet 10^{-3}$ |
| $Mg_{0.05}La_{0.95}Co_{10}Ni_2Si$<br>(6Q/N57 MLCNI13) | 0.89<br>0.04<br>0.07 | **$Mg_{0.2}La_7Co_{71}Ni_{13}Si_8$**<br>$Mg_{0.2}La_{14}Co_{46}Ni_{33}Si_7$<br>Co | $NaZn_{13}$<br>$CaCu_5$<br>Mg | 1.1300(2)<br>0.5014(6)<br>0.2507(9) | <br>0.4000(5)<br>0.4070(9) | 14.1 | 0.66 | 18.0 | 1.62 | $5.0 \bullet 10^{-3}$ |
| $Ce_{0.05}La_{0.95}Co_{10}Fe_2Si$<br>(6Q/N49 CLCFI13) | 0.94<br>0.06 | **$Ce_{0.2}La_7Co_{71}Fe_{13}Si_8$** | $NaZn_{13}$<br>W | 1.1323(2)<br>0.2835(5) | | 13.8 | 0.742 | 15.8 | 1.6 | $4.9 \bullet 10^{-3}$ |
| $Y_{0.05}La_{0.95}Co_{10}Fe_2Si$<br>(6Q/N50 YLCFI13) | 0.98<br>0.02 | **$Y_{0.1}La_7Co_{71}Fe_{14}Si_8$**<br>$Y_3La_5Co_{71}Fe_{15}Si_6$ | $NaZn_{13}$<br>W | 1.1323(2)<br>0.2835(5) | | 12.7 | 0.74 | 16.0 | 1.62 | $4.1 \bullet 10^{-3}$ |
| $Mg_{0.05}La_{0.95}Co_{10}Fe_2Si$<br>(6Q/N51 MLCFI13) | 0.90<br>0.10 | **$Mg_{0.2}La_7Co_{71}Fe_{14}Si_8$**<br>x | $NaZn_{13}$<br>Mg | 1.1331(2)<br>0.2649(2) | <br>0.4097(2) | 14.6 | 0.642 | 18.0 | 1.58 | $5.5 \bullet 10^{-3}$ |
| $Ce_{0.05}La_{0.95}Co_{10}Mn_2Si$<br>(6Q/N52 CLCMI13) | 0.93<br>0.06<br>0.02 | **$Ce_{0.2}La_7Co_{71}Mn_{14}Si_8$**<br>$La_7Co_{69}Mn_{22}$<br>$Co_{74}Mn_{24}$ | $NaZn_{13}$<br>$ThMn_{12}$<br>W | 1.1369(2)<br>0.8447(5)<br>0.2806(9) | <br>0.4252(3) | 4.25 | 1.14 | 11.6 | 1.8 | $4.1 \bullet 10^{-3}$ |
| $Y_{0.05}La_{0.95}Co_{10}Mn_2Si$<br>(6Q/N53 YLCMI13) | 0.94<br>0.02 | **$Y_{0.1}La_7Co_{71}Mn_{14}Si_8$**<br>$Y_3La_6Co_{65}Mn_{20}Si_7$<br>$Y_3La_6Co_{68}Mn_{22}$ | $NaZn_{13}$<br>$ThMn_{12}$<br>$CaCu_5$ | 1.1379(2)<br>0.4735(7) | <br>0.4216(5) | 4.35 | 1.16 | 11.1 | 1.76 | $4.4 \bullet 10^{-4}$ |
| $Mg_{0.05}La_{0.95}Co_{10}Mn_2Si$<br>(6Q/N54 MLCMI13) | 0.96<br>0.02<br>0.02 | **$Mg_{0.2}La_8Co_{73}Mn_{10}Si_8$**<br>$Co_{76}Mn_{22}$<br>x | $NaZn_{13}$<br>W<br>$ThMn_{12}$ | 1.1360(2)<br>0.2896(8)<br>0.8459(9) | <br><br>0.4290(5) | 4.62 | 1.13 | 12.2 | 1.88 | $4.6 \bullet 10^{-4}$ |
| **$CeCu_6MnAl_6$**<br>(6Q/N99 CPMA13) | 1.00 | $Ce_7Cu_{48}Mn_2Al_{43}$<br>$Ce_7Cu_{45}Mn_5Al_{43}$ | $NaZn_{13}$ | 1.1815(2) | | 11.0 | $2.2^{SC}$ | 10.1 | 3.03 | $1.6 \bullet 10^{-3}$ |

Table 3

Composition, cell parameters and physical properties of $R_xR'_{1-x}T_yNi_{1-y}Sb_zX_{1-z}$ solid solutions (R, R' = Y, Ce, Yb; T = Transition metal)

| **Alloy**<br>(sample) | **Mass fraction** | **Phase** | **Structure** | **a** | **c** | **S** | **ρ** | **κ** | **WF** | **ZT** |
|---|---|---|---|---|---|---|---|---|---|---|
| | | CeNiSb | $InNi_2$ | 0.4395 | 0.8258 | | | | | |
| $Ce_{0.75}Y_{0.25}NiSb$<br>(5Q/N6 CYNSb1C) | 0.90<br>0.10 | **$Ce_{26}Y_{10}Ni_{32}Sb_{32}$**<br>$Ce_{29}Y_{14}Ni_{19}Sb_{39}$<br>$Ce_{17}Y_{11}Ni_{35}Sb_{37}$ | $InNi_2$<br>MgAgAs<br>$CaBe_2Ge_2$ | 0.4384(1)<br>0.6402(2) | 0.8039(1) | 10.0 | 5.4 | 6.2 | 4.57 | $9.0\bullet10^{-4}$ |
| $Ce_{0.5}Y_{0.5}NiSb$<br>(3Q/N97 CYNSb1) | 0.93<br>0.07 | **$Ce_{18}Y_{17}Ni_{32}Sb_{33}$**<br>$Ce_{33}Y_{17}Sb_{50}$ | $InNi_2$<br>NaCl | 0.4374(1)<br>0.6375(3) | 0.7792(1) | -53.2 | 1.66 | 11.0 | 2.49 | **0.0465** |
| $Ce_{0.5}Y_{0.5}NiSb^{HP}$<br>(8Q/142 40T) | ~0.70 | **$Ce_{25}Y_{22}Ni_{21}Sb_{28}$**<br>**$Ce_{25}Y_{22}Ni_{13}Sb_{41}$**<br>$Ce_2Y_{22}Ni_{77}$<br>$Ce_{34}Y_{13}Ni_{10}Sb_{43}$ | MgAgAs<br>MgAgAs<br>$CaCu_5$<br>$Th_3P_4$ | 0.6343(2) | | 4.02 | 3.07 | 6.81 | 2.84 | $2.3\bullet10^{-4}$ |
| $Ce_{0.5}Y_{0.5}NiSb$<br>(5Q/N5 CYNSb1B') | 0.60<br>0.27<br>0.13 | **$Ce_{17}Y_{17}Ni_{33}Sb_{33}$**<br>$Ce_{18}Y_{22}Ni_{22}Sb_{38}$<br>$Ce_5Y_{14}Ni_{80}$ | $InNi_2$<br>MgAgAs<br>$CaCu_5$ | 0.4380(1)<br>0.6368(2)<br>0.5189(6) | 0.7789(2)<br><br>0.3788(3) | -10.5 | 6.72 | 6.3 | 5.78 | $7.8\bullet10^{-4}$ |
| $Ce_{0.5}Y_{0.5}NiSb^{HP}$<br>(6Q/N60 CYNSb1X - 23A) | 0.98<br>0.02 | **$Ce_{18}Y_{18}Ni_{32}Sb_{32}$**<br>$Ce_{14}Y_{27}Ni_{20}Sb_{39}$ | $InNi_2$<br>MgAgAs | 0.4372(2)<br>0.6335(7) | 0.7821(2) | -22.8 | 48.25 | 2.45 | 16.1 | $1.3\bullet10^{-3}$ |
| $Ce_{0.25}Y_{0.75}NiSb$<br>(4Q/N147 CYNSb1B) | 0.70<br>0.30 | **$Ce_{10}Y_{26}Ni_{32}Sb_{32}$**<br>$Ce_{12}Y_{13}Ni_{42}Sb_{35}$ | $InNi_2$<br>MgAgAs<br>$CaBe_2Ge_2$ | 0.4358(2)<br>0.6328(2) | 0.7626(2) | -52.6 | $9.99^{SC}$ | 5.7 | 7.77 | 0.015 |
| $Ce_{0.1}Y_{0.9}NiSb$<br>(6Q/N64 CYNSb4) | | **$Ce_5Y_{31}Ni_{30}Sb_{35}$**<br>$Ce_4Y_{21}Ni_{48}Sb_{28}$ | MgAgAs<br>x | 0.6307(1) | | -19.0 | $7.9^{SC}$ | 6.25 | 6.74 | $2.2\bullet10^{-3}$ |
| $Yb_{0.25}Y_{0.75}NiSb$<br>(4Q/N148 YbNSb2) | 0.69<br>0.31 | **$Yb_6Y_{31}Ni_{29}Sb_{34}$**<br>$Yb_7Y_{17}Ni_{41}Sb_{51}$ | MgAgAs<br>$CaBe_2Ge_2$ | 0.6276(1) | | -26.0 | 3.36 | 8.7 | 3.99 | $6.9\bullet10^{-3}$ |
| $Yb_{0.5}Y_{0.5}NiSb$<br>(4Q/N149 YbYNSb1) | 0.64<br>0.35 | **$Yb_{12}Y_{28}Ni_{24}Sb_{37}$**<br>$Yb_{11}Y_{12}Ni_{41}Sb_{37}$ | MgAgAs<br>$CaBe_2Ge_2$ | 0.6255(1)<br>0.4215(1) | <br>0.9682(3) | -5.0 | 1.88 | 9.5 | 2.44 | $4.2\bullet10^{-4}$ |
| $YbCu_2Sb_2$<br>(4Q/N132 YbPSb122A) | 0.77<br>0.14<br>0.09 | $Yb_{26}Cu_{38}Sb_{38}$<br>$Cu_{67}Sb_{33}$<br>Cu | $AlB_2$<br>NiAs<br>Cu | 0.4450(1)<br>0.4000(2)<br>0.3611(2) | 0.3989(1)<br>0.5153(2) | -1.6 | 1.4 | 13.0 | 2.48 | $4.2\bullet10^{-5}$ |
| $Ce_{0.5}Y_{0.5}Ni_{0.99}Cu_{0.01}$ Sb<br>(6Q/N63 CYNPSb4) | 0.71<br>0.29 | **$Ce_{16}Y_{18}Ni_{33}Cu_1Sb_{32}$**<br>x<br>$Ce_{14}Y_7Ni_{48}Cu_2Sb_{30}$ | $InNi_2$<br>MgAgAs<br>x | 0.4377(1)<br>0.6371(2) | 0.7816(2) | -24.7 | 4.7 | 6.5 | 4.17 | $6.0\bullet10^{-3}$ |
| $Ce_{0.5}Y_{0.5}Ni_{0.95}Cu_{0.05}$ Sb<br>(5Q/N8 YCNPSb1) | 0.89<br>0.11 | **$Ce_{17}Y_{18}Ni_{32}Cu_2Ni_{32}Sb_{33}$**<br>$Ce_{17}Y_{21}Ni_{26}Sb_{36}$<br>$Ce_{15}Y_8Ni_{20}Cu_4Sb_{33}$ | $InNi_2$<br>MgAgAs<br>$ZrCuSi_2$ | 0.4370(1)<br>0.6364(3) | 0.7814(2) | -18.1 | 5.17 | 0.42 | 0.30 | 0.045 |
| $Ce_{0.5}La_{0.5}NiSb$<br>(5Q/N7 LCNSb1) | 0.84<br>0.06<br>0.10 | **$La_{17}Ce_{17}Ni_{32}Sb_{35}$**<br>$La_{23}Ce_{22}Ni_{10}Sb_{36}$<br>$La_9Ce_{12}Ni_{43}Sb_{36}$ | $InNi_2$<br>MgAgAs<br>$ZrCuSi_2$ | 0.4394(1)<br>0.6480(2)<br>0.4334(2) | 0.8307(2)<br><br>1.0113(7) | -3.85 | 7.1 | 4.48 | 4.34 | $1.4\bullet10^{-4}$ |
| $Ce_{0.5}Y_{0.5}NiSb_{0.5}Sn_{0.5}$<br>(5Q/N15 CYNSbSn1) | | **$Ce_{16}Y_{18}Ni_{33}Sb_{13}Sn_{20}$**<br>$Ce_{19}Y_{20}Ni_{24}Sb_{34}Sn_2$<br>$Ce_{10}Y_6Ni_{65}Sn_{19}$ | $InNi_2$<br>MgAgAs<br>x | 0.4428(2)<br>0.6371(2) | 0.7389(4) | -6.6 | 3.0 | 7.5 | 3.07 | $5.8\bullet10^{-4}$ |
| $Ce_{0.5}Y_{0.5}NiSb_{0.5}Bi_{0.5}$<br>(5Q/N4 CYNSbBi1A) | 0.62<br>0.38 | **$Ce_{20}Y_{21}Ni_{19}Sb_{22}Bi_{18}$**<br>$Ce_6Y_{15}Ni_{74}Sb_5Bi_2$ | MgAgAs<br>$CaCu_5$ | 0.6401(2)<br>0.4896(3) | <br>0.3975(2) | 6.9 | 1.67 | 10.8 | 2.46 | $7.9\bullet10^{-4}$ |
| $Ce_{0.5}Y_{0.5}NiSb_{0.8}Bi_{0.2}$<br>(6Q/62 CYNSbBi4) | 0.77<br>0.23 | **$Ce_{18}Y_{18}Ni_{31}Si_{32}Bi_3$**<br>$Ce_{26}Y_{26}Sb_{46}Bi_4$<br>$Ce_{15}Y_8Ni_{44}Sb_{30}Bi_3$ | $InNi_2$<br>NaCl<br>x | 0.4377(2)<br>0.6320(2) | 0.7831(2) | -36.8 | 2.42 | 7.7 | 2.54 | 0.022 |

Table 4

Composition, cell parameters and physical properties of alloys of the Mg-Ce-Ni and Mg-Sm-Ni systems

| **Alloys**<br>(sample) | **Mass fraction** | **Phase** | **Structure** | **a** | **c** | **S** | ρ | κ | **WF** | **ZT** |
|---|---|---|---|---|---|---|---|---|---|---|
| $Ce_2MgNi_2$ $^{HP}$<br>(7Q/112 7T) | 1.00 | **$Ce_{41}Mg_{20}Ni_{39}$** | $Mo_2FeB_2$ | 0.7579(3) | 0.3760(1) | -22.5 | 0.46 | 8.6 | 0.54 | 0.038 |
| $CeYMgNi_2$ $^{HP}$<br>(8Q/132 37T) | ~0.80 | **$Ce_{15}Y_{25}Mg_{20}Ni_{40}$**<br>$Ce_{13}Y_6Mg_{15}Ni_{66}$<br>$Ce_{25}Y_{23}Mg_3Ni_{49}$<br>$Ce_{19}Y_3Mg_{53}Ni_{25}$ | $Mo_2FeB_2$<br>$AuBe_5$<br>CrB<br>x | 0.7490(3) | 0.3700(2) | -16.85 | 0.91 | 13.3 | 1.65 | 0.007 |
| $Sm_2MgNi_2$ $^{HP}$<br>(8Q/133 34T) | ~0.90 | **$Sm_{40}Mg_{21}Ni_{39}$**<br>$Sm_{46}Mg_{10}Ni_{44}$<br>$Sm_{17}Mg_{64}Ni_{19}$ | $Mo_2FeB_2$<br>x<br>x | 0.7470(4) | 0.3758(3) | 2.37 | 0.80 | 18.9 | 2.06 | $1.1•10^{-4}$ |
| $SmYMgNi_2$ $^{HP}$<br>(8Q/135 38T) | ~0.70 | **$Sm_{31}Y_9Mg_{21}Ni_{38}$**<br>$Sm_{29}Y_6Mg_2Ni_{64}$<br>$Y_{80}Ni_{20}$<br>$Sm_{13}Y_2Mg_{63}Ni_{23}$ | $Mo_2FeB_2$<br>$AuBe_5$<br>x<br>x | 0.7436(5) | 0.3741(3) | -0.42 | 0.76 | 16.1 | 1.66 | $4.2•10^{-6}$ |
| $Ce_{1.5}Mg_{1.5}Ni_9$<br>(4Q/N160 CMgN1A) | 0.56<br>0.07<br>0.37 | $Ce_{19}Mg_{13}Ni_{68}$<br>$Ce_{19}Ni_{80}$<br>$Ce_{19}Ni_{80}$ | $AuBe_5$<br>$Ce_2Ni_7$<br>$CaCu_5$ | 0.7005(2)<br>0.4891(3)<br>0.4892(15) | <br>2.391(1)<br>0.4002(8) | -24.0 | 0.51 | 16 | 1.1 | 0.021 |
| $Ce_{2.5}Mg_{0.5}Ni_9$<br>(3Q/N91 MgN3) | 0.42<br>0.05<br>0.50 | $Ce_{19}Mg_{12}Ni_{68}$<br>$Ce_{20}Ni_{80}$<br>$Ce_{20}Ni_{80}$ | $AuBe_5$<br>$Ce_2Ni_7$<br>$CaCu_5$ | 0.6992(2)<br>0.4950(8)<br>0.4887(3) | <br>2.423(3)<br>0.3998(2) | -18.0 | 0.65 | 33.0 | 2.9 | $4.5•10^{-3}$ |
| $Ce_{19}Mg_{12}Ni_{69}$<br>(4Q/N161 CMgN1B) | | $Ce_{19}Mg_{13}Ni_{67}$<br>$Ce_{15}Mg_{10}Ni_{75}$ | $AuBe_5$<br>x | 0.7064(5) | | 27.2 | 2.63 | 12.5 | 4.5 | $6.8•10^{-3}$ |
| $Ce_{0.5}Mg_{0.5}Ni_2$ $^{HP}$<br>(6Q/N73 17E) | | $Ce_{18}Mg_{15}Ni_{69}$<br>$Ce_{17}Ni_{83}$ | $AuBe_5$<br>$CaCu_5$ | 0.70075(6) | | -29.4 | 0.88 | 14.0 | 1.68 | 0.021 |
| $Ce_{0.5}Mg_{0.5}Ni_2$<br>(6Q/N70 CMgN22) | | $Ce_{18}Mg_{13}Ni_{69}$<br>$Ce_{18}Mg_6Ni_{76}$<br>$Ce_{17}Ni_{80}$ | $AuBe_5$<br>$PuNi_3$<br>$CaCu_5$ | 0.70085(9) | | -33.6 | 1.00 | 12.3 | 1.68 | 0.028 |
| $Ce_{0.5}Mg_{0.5}Ni_2$ $^{HP}$<br>(6Q/N71 CMgN22 - 25A) | | $Ce_{18}Mg_{11}Ni_{70}$<br>$Ce_{20}Mg_4Ni_{76}$<br>$Ce_{19}Mg_2Ni_{79}$<br>$Ce_{26}Mg_{19}Ni_{55}$ | $AuBe_5$<br>$PuNi_3$<br><br>x | 0.69961(9) | | -35.6 | 1.07 | 11.7 | 1.71 | 0.030 |
| $Ce_{0.5}Mg_{0.5}Ni_2$ $^{HP}$<br>(6Q/N75 18H - 13B) | | $Ce_{21}Mg_{12}Ni_{67}$<br>$Ce_{48}Mg_{50}$<br>$Mg_{33}Ni_{67}$<br>$Ce_{17}Ni_{82}$ | $AuBe_5$<br>CsCl<br>$MgNi_2$<br>$CaCu_5$ | 0.7032(2) | | -4.5 | 1.44 | 10.0 | 1.96 | $4.2•10^{-4}$ |
| $Ce_{0.5}Y_{0.5}MgNi_4$<br>(8Q/N138 43T) | ~0.80 | $Ce_8Y_8Mg_{18}Ni_{66}$<br>$Ce_2Y_2Mg_{29}Ni_{68}$<br>$Ce_6Y_7Mg_{13}Ni_{74}$ | $AuBe_5$<br>$MgNi_2$<br>$CeNi_3$ | 0.7010(2) | | -10.2 | 1.12 | 10.95 | 1.66 | $2.6•10^{-3}$ |
| $Ce_{0.5}Mg_{0.5}CoNi$<br>(7Q/108 3T) | ~1.0 | **$Ce_{19}Mg_{16}Ni_{47}Co_{19}$**<br>$Co_{99}$<br>$Ce_{36}Co_{63}$<br>$Ce_{10}Mg_{74}Ni_{14}$ | $AuBe_5$<br>Mg<br>$MgCu_2$<br>mf | 0.7050(1) | | 4.45 | 0.76 | 16.0 | 1.66 | $4.9•10^{-4}$ |
| $CeMg_2Ni_9$ $^{HP}$<br>(6Q/N78 15E) | 0.70<br>0.20<br>0.10 | $Ce_9Mg_{16}Ni_{76}$<br>$Ce_{15}Mg_2Ni_{84}$<br>$Ni_{99}$<br>$Ce_7Mg_{28}Ni_{65}$ | $PuNi_3$<br>$CaCu_5$<br>Cu<br>x | 0.4861(1)<br>0.4872(2)<br>0.3534(3) | 2.3845(4)<br>0.4019(1) | 3.68 | 0.62 | 17.4 | 1.47 | $3.8•10^{-4}$ |
| $CeMg_2Ni_9$ $^{HP}$<br>(6Q/N79 19H-11B) | 0.38<br>0.52<br>0.10 | $Mg_{24}Ni_{75}$<br>$Ce_{17}Ni_{83}$<br>$Ni_{99}$ | $MgNi_2$<br>$CaCu_5$<br>Cu | 0.4824(2)<br>0.4886(1)<br>0.3536(3) | 1.5829(8)<br>0.4002(1) | 6.3 | 0.54 | 21.5 | 1.58 | $1.0•10^{-3}$ |

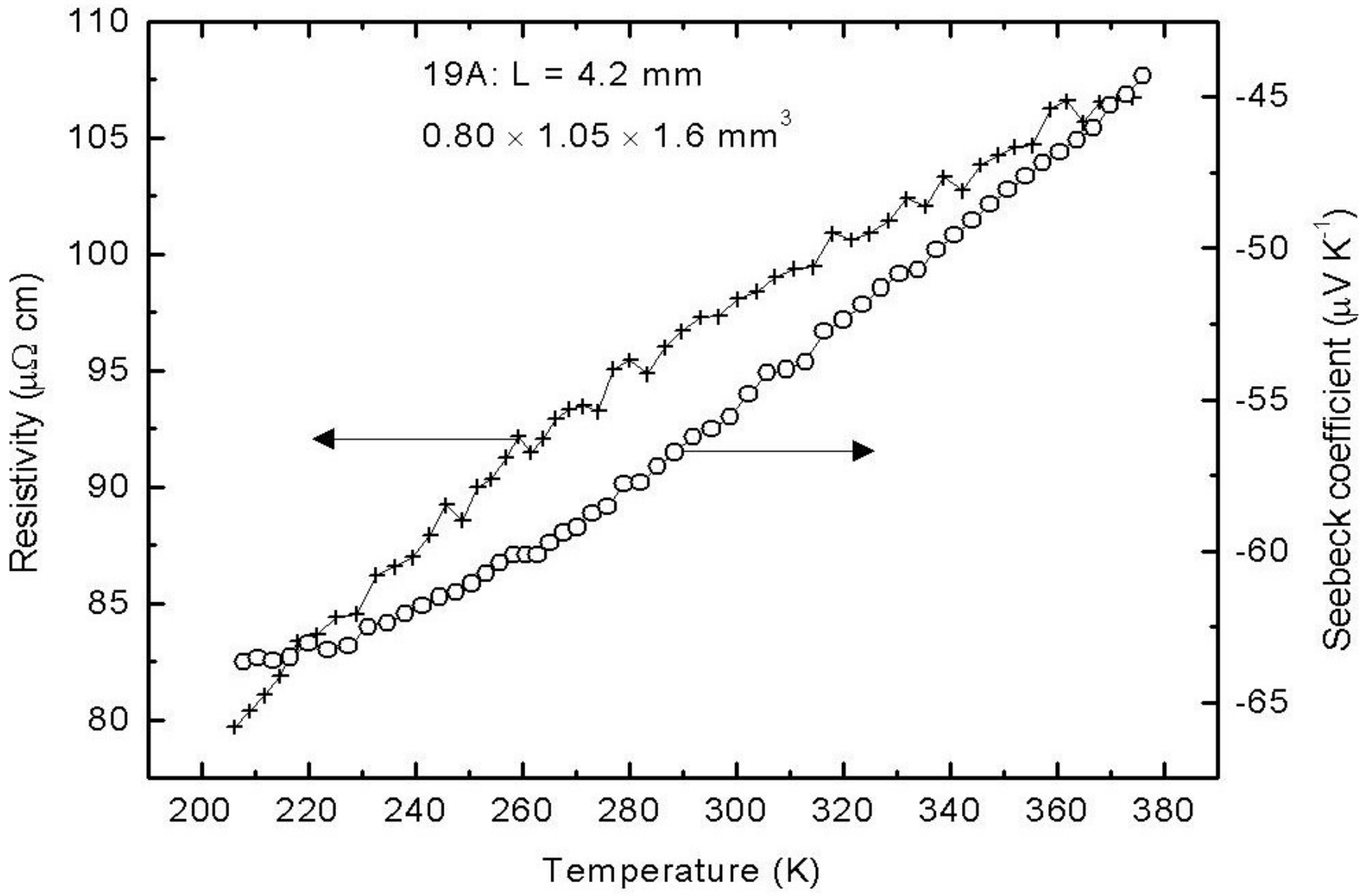

**(a)**

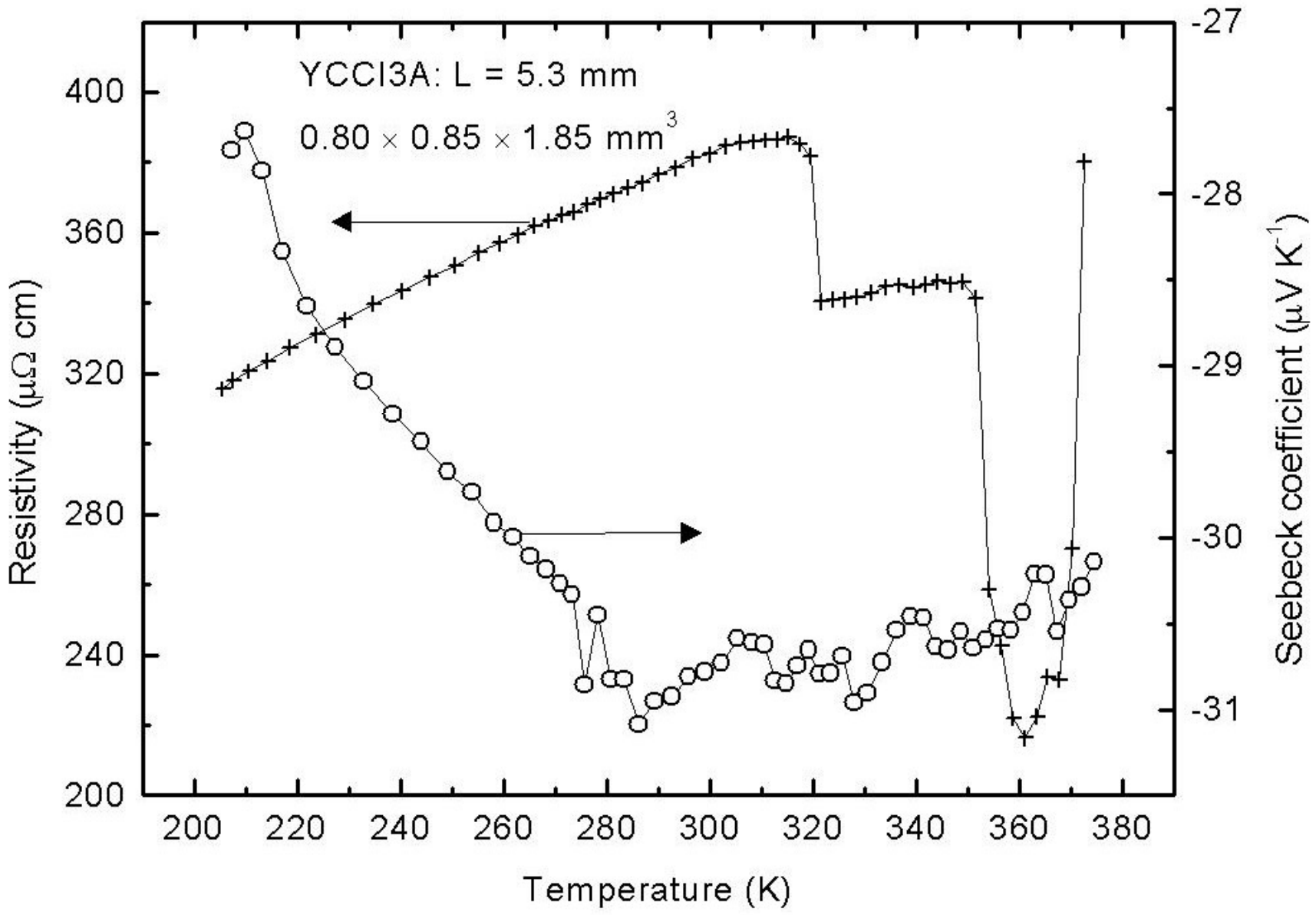

**(b)**

Fig.1. Resistivity and Seebeck coefficient vs. temperature for $CeSi_2$ (a) and $Ce_{0.5}Y_{0.5}CoSi_3$ (b) alloys

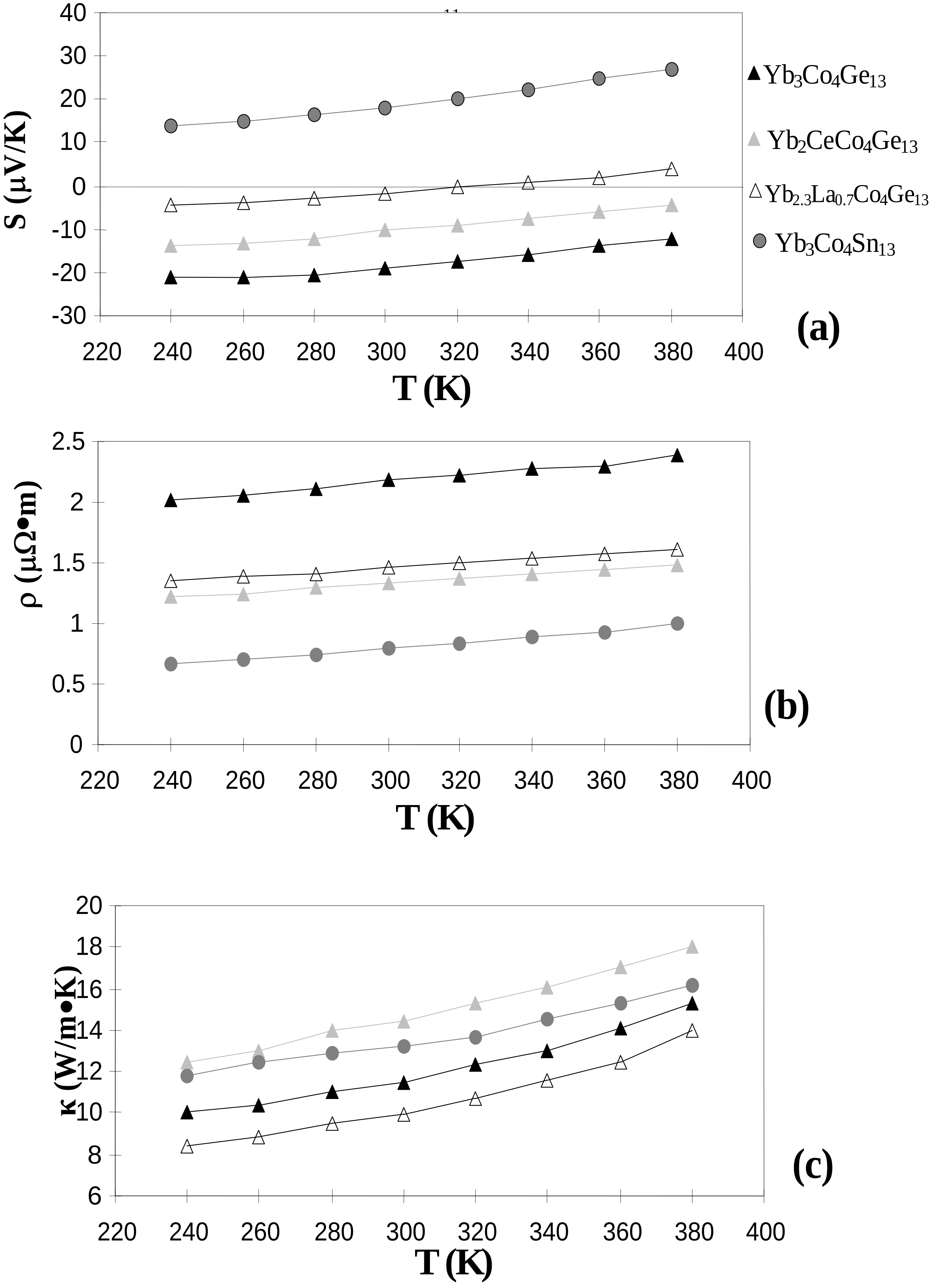

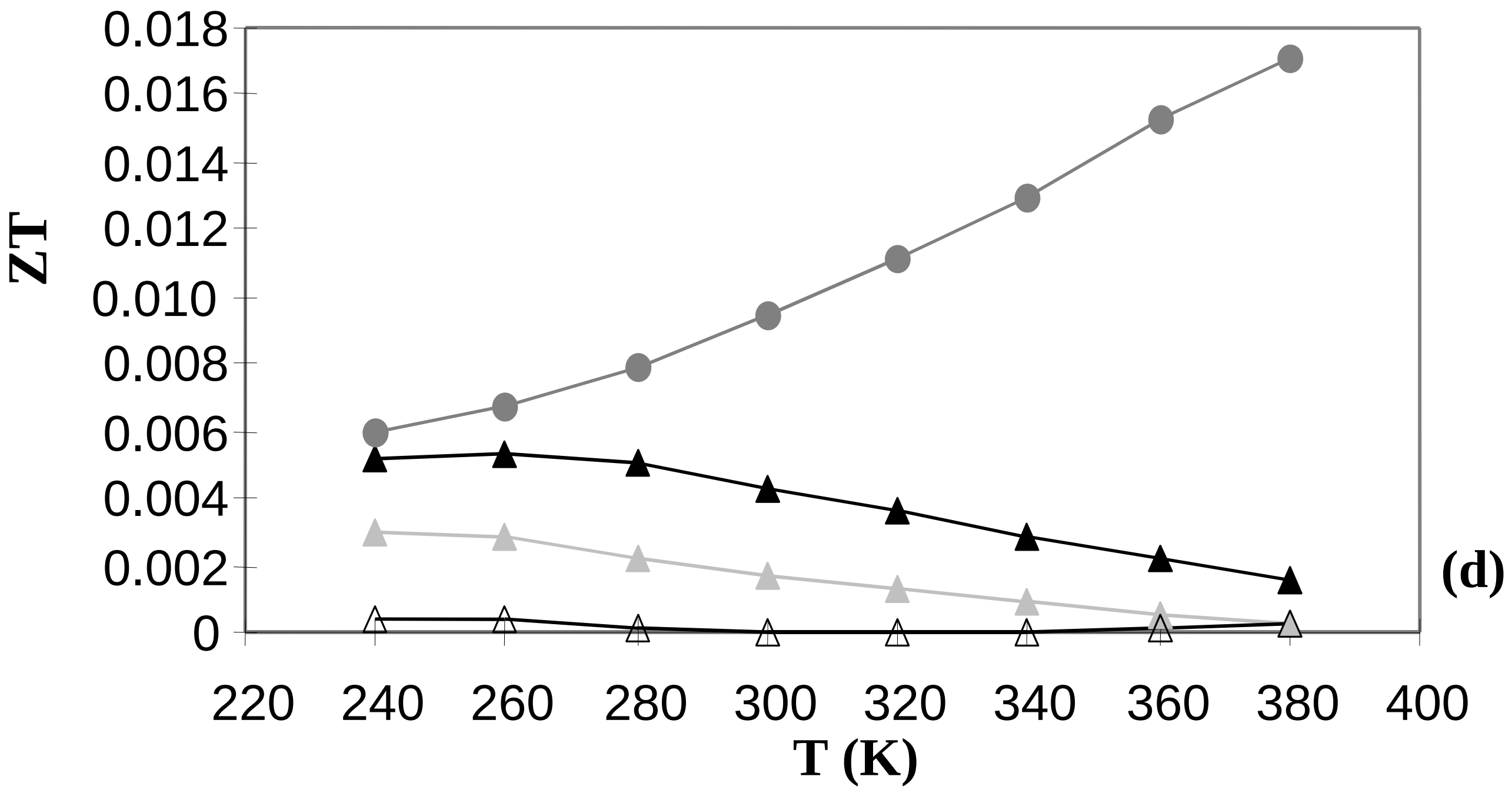

**Fig. 2**. Seebeck coefficient *S*, electric resistance ρ, thermal conductivity κ and *ZT* parameter vs. temperature for $Yb_3Co_4Ge_{13}$, $Yb_2CeCo_4Ge_{13}$, $Yb_{2.3}La_{0.7}Co_4Ge_{13}$ and $Yb_3Co_4Sn_{13}$ compounds